# Stimulated Rayleigh Scattering Enhanced by a Longitudinal Plasma Mode in a Periodically Driven Dirac Semimetal $Cd_3As_2$

Yuta Murotani[1†*], Natsuki Kanda[1,2†*], Tatsuhiko N. Ikeda[1],
Takuya Matsuda[1], Manik Goyal[3], Jun Yoshinobu[1],
Yohei Kobayashi[1], Susanne Stemmer[3], and Ryusuke Matsunaga[1,2*]

[1]*The Institute for Solid State Physics, The University of Tokyo, Kashiwa, Chiba 277-8581, Japan*
[2]*PRESTO, Japan Science and Technology Agency, 4-1-8 Honcho Kawaguchi, Saitama 332-0012, Japan*
[3]*Materials Department, University of California, Santa Barbara, California 93106-5050, USA*
[†]*These authors contributed equally to this work.*
*e-mail: murotani@issp.u-tokyo.ac.jp, n-kanda@issp.u-tokyo.ac.jp, matsunaga@issp.u-tokyo.ac.jp

**Abstract**

Using broadband (12-45 THz) multi-terahertz spectroscopy, we show that stimulated Rayleigh scattering dominates the transient optical conductivity of cadmium arsenide, a Dirac semimetal, under an optical driving field at 30 THz. The characteristic dispersive lineshape with net optical gain is accounted for by optical transitions between light-induced Floquet subbands, strikingly enhanced by the longitudinal plasma mode. Stimulated Rayleigh scattering with an unprecedentedly large refractive index change may pave the way for slow light generation in conductive solids at room temperature.

**Main text**

Light has opened various ways to reach interesting nonequilibrium phases of matter, such as light-induced superconductivity [1,2], charge density wave [3], and excitonic insulator [4]. The emerging field of Floquet engineering is accelerating new discoveries through the versatility of periodic driving to modify material properties [5,6]. Examples include control of band topology [7-10] and of excitonic correlations [11,12]. Floquet engineering is also interesting from the viewpoint of nonlinear optics. Historically, the concept of photon-dressed states has provided an indispensable basis to understand the nonlinear optical response of discrete level systems [13]. Modern interest in Floquet engineering has extended the idea of dressed states to continuous bands in solids, revealing new aspects of nonlinear optics, e.g., in terms of topology [14]. It is thus natural to expect novel optical phenomena to emerge from light-induced Floquet states.

Despite remarkable progress in theory, experimental exploration of Floquet states is still limited. Time- and angle-resolved photoemission spectroscopy succeeded in directly observing electron population in photon-dressed Floquet-Bloch bands on a surface of a light-driven topological insulator [15,16]. Ultrafast transport measurement has recently demonstrated that irradiation by circularly polarized light transforms graphene into a Floquet topological insulator [7,17], which partly contributes to anomalous Hall effect [18]. Manifestations of the light-induced Floquet states in the optical response itself, however, remain unclear. Little has been known about fundamental optical properties of Floquet states in solids, except for the well-known ac Stark effect of discrete levels. Cadmium arsenide ($Cd_3As_2$), a three-dimensional Dirac semimetal, is an ideal material to investigate this problem, because it combines high-mobility carriers, a small scattering rate, and low-energy interband transitions [19], which allow for coherent dynamics with suppressed dissipation and laser heating. Moreover, $Cd_3As_2$ exhibits large optical nonlinearity in a broad frequency region ranging from terahertz to visible [20-25], which makes it a promising platform to search for novel functionality in nonlinear optics and optoelectronics from the perspective of Floquet engineering.

Figure 1(a) shows the band structure of $Cd_3As_2$. Two Dirac nodes lie on the $k_z$ axis, which allow low-energy interband transitions [19,26]. The valence and conduction bands are expected to form Floquet states upon periodic driving by a light field, as shown in Fig. 1(b). To explore the spectroscopic signature and optical functionality of the Floquet states, we measure transient optical conductivity of an epitaxially grown, (112)-oriented, 140 nm-thick $Cd_3As_2$ thin film on a CdTe substrate [27], exposed to an intense multi-terahertz electromagnetic pulse at room temperature. Figure 1(c) depicts the experimental setup. Our sample is unintentionally electron-doped so that the Fermi level is shifted to 58 meV above the Dirac nodes [25]. Despite the anisotropy in the low-energy band structure, the linear response in the infrared region is almost isotropic because of the quasi-cubic nature of the structural units that make up the unit cell [36,37]. Figure 1(d) shows the optical conductivity of the sample in equilibrium. It can be decomposed into the low-frequency (<15 THz) intraband and high-frequency (>15 THz) interband contributions, by taking account of the low-frequency data outside the panel [25,38]. The narrowband pump pulse drives the interband transitions with a tunable frequency from 16 to 40 THz (66-165 meV in energy, 8-19 μm in wavelength) and with a variable bandwidth, while the probe pulse covers a broad frequency range from 12 to 45 THz (50-186 meV, 7-25 μm) with a duration of 30 fs. The probe pulse after transmitting the sample is spatially separated from the pump pulse and is detected by electro-optic sampling to obtain response functions depending on the pump-probe delay time $\Delta t$ [27].

Figure 2(a) shows the transient optical conductivity measured by probe pulses polarized in the same direction as the pump, tuned to 29.4 THz. During the pump irradiation, a photoinduced absorption (blue) appears just below the pump frequency, while an opposite change (red) occurs on the higher-frequency side. The resulting dispersive lineshape is clearly seen in Fig. 2(b), which plots the optical conductivity at several delay times. This characteristic behavior is distinct both from spectral hole burning [39] and from photon-assisted absorption bands [40], the two scenarios that have been theoretically considered so far. Note that net optical gain ($\sigma_1 < 0$) develops from the suppressed absorption at around the maximum pump-probe overlap ($\Delta t \simeq 0$ ps). The dispersive structure vanishes after the pump pulse leaves the sample, as visualized in Fig. 2(d). Upon changing the pump fluence, positions of the peak and the dip stay almost constant, as shown in Fig. 2(e). We also plot in the same figure the fluence dependence of the peak and dip values along with the equilibrium values at 28.2 and 31.3 THz (open triangles). In the weak excitation limit (<0.1 mJ/cm$^2$), both the peak and the dip grow linearly with the pump fluence, indicating a perturbative origin of the signal. We found that no dispersive signal appears when the probe is polarized perpendicularly to the pump [27], implying a coherent nature of the involved processes. In addition, Fig. 2(c) verifies that the position of the dispersive structure follows the center frequency of the pump, excluding the possibility that the signal could arise from some special points in the band structure or specific phonon modes.

In the case of semiconductors, it is known that a dispersive absorption change appears in the early stage of photoexcitation as a result of excitonic effect [41,42]. In this mechanism, however, the absorption peak should lie on the higher energy side of the pump photon energy, which is opposite to the behavior observed here. Thus, excitonic effects are of minor importance in Cd$_3$As$_2$, consistent with recent predictions [43].

From a phenomenological point of view, the dispersive absorption change in Cd$_3$As$_2$ can be understood in terms of stimulated Rayleigh scattering (SRLS). Suppose that application of the optical field primarily changes the real part of the refractive index. When the pump and probe beams spatially overlap, their interference creates a transient grating, which diffracts the pump beam into two directions; one is a new direction often studied in four-wave mixing experiments, and the other the propagation direction of the probe, as shown in Fig. 3(e). The latter effect suppresses or enhances absorption of the probe beam depending on the phase of the diffracted wave. In case of a negative refractive index change, this process results in a photoinduced absorption (emission) for a probe frequency slightly lower (higher) than the pump, as seen in Fig. 3(a). This mechanism accounts for our experimental results, because interband excitation actually reduces the refractive index through a blueshift of the longitudinal plasma mode initially located at 10 THz [25,27]. The blueshift is associated with increased density of charge carriers, since the squared plasma frequency is proportional to the carrier density. In nonlinear optics, light scattering by light-induced density fluctuations of gases and crystals is known as SRLS [13]. Therefore, the process described above also belongs to SRLS, which utilizes the collective plasma oscillation of charge carriers as a novel source of it. We note that this mechanism of SRLS is distinct from the conventional ones not only qualitatively – in its origin – but also quantitatively. The collective nature of the plasma mode enables a large refractive index change more than 1 [27], which far exceeds the known cases and thus leads to unprecedentedly strong SRLS. It is interesting that metallic response of solids with the plasma mode significantly enhances the coherent light-matter interaction. We will discuss a possible application of such a large refractive index change later. In the phenomenological model presented above, the separation between the peak and the dip decreases for increasing $\Delta t$ as shown in Fig. 3(b), consistent with the experimental result in Fig. 2(a). Such a narrowing is explained by the detection scheme in our experiment [27].

We next consider the quantum mechanical aspect of this phenomenon and discuss its connection to the Floquet states. In Fig. 3(c), we plot the transient optical conductivity calculated by an effective two-band model for the low-energy band structure [27]. One can clearly recognize a dispersive lineshape. Knowledge of two-level systems helps us to interpret this result using a level diagram. In two-level systems, the well-known ac Stark effect is accompanied by a dispersive structure at the pump frequency, also called SRLS [13,27]. It originates from transitions between dressed states in resonance with the driving field. Extending this understanding to the continuous bands, SRLS in $Cd_3As_2$ is attributed to transitions between the Floquet subbands resonant to the pump frequency, as schematically shown in Fig. 3(f). Relatively small scattering rates in $Cd_3As_2$ justify such a Floquet state picture. A closer look at its origin, however, reveals the difference of light-matter interaction responsible for SRLS in $Cd_3As_2$ and in two-level systems. In the latter, a usual coupling between electric dipole moments and the electric field, also called paramagnetic coupling, induces relatively weak SRLS, with a sign depending on detuning [13]. As a result, SRLS in two-level systems tends to be cancelled out when integrated over continuous bands, leaving a spectral hole stemming from the ac Stark effect and Pauli blocking [39]. This consequence can be seen in the blue curve in Fig. 3(d), which plots the contribution from the paramagnetic coupling only. The dispersive structure in the total optical conductivity arises from a second-order or diamagnetic coupling with the electric field, which yields the red curve in Fig. 3(d) showing good agreement with the experimental result. This coupling causes a light-induced shift of the screened plasma frequency, so that the microscopic theory also supports the phenomenological picture presented above. In Supplemental Material, we derive the macroscopic model by analyzing the diamagnetic current in the microscopic model [27]. The derivation tells us that an intermediate frequency between intraband and interband transitions is preferable, because SRLS in this case requires combination of injection and acceleration of photocarriers. These findings renew the prospect of Floquet engineering for optical properties of matter, because importance of the diamagnetic coupling has not been recognized so far. Since the above discussion does not rely on details of the band structure, SRLS is expected to occur in general semimetals and narrow-gap semiconductors with low-energy interband transitions.

Finally, from a perspective of Floquet engineering of optical functionality, we discuss the possibility of slow light generation in $Cd_3As_2$. Consistent with a general property of SRLS [13], the dispersive structure in transient optical conductivity can be narrowed by reducing the pump bandwidth, as shown in Fig. 4(a). Such a narrow structure in absorption is necessarily accompanied with a rapid variation in the refractive index $n$ with frequency $f$, so that the group refractive index $n_g = n + f(dn/df)$ may become large. The resultant slowing down of an optical wave packet is known as slow light generation [13,44-50]. In the present experiment, we directly evaluate the broadband refractive index as a complex quantity. The top panel in Fig. 4(b) shows that a narrow dip in the refractive index develops at, e.g., $\Delta t = -0.48$ ps, leading to a group refractive index as large as 40 at 30 THz (bottom panel in Fig. 4(b)). This corresponds to 40 times deceleration of a wave packet, free from dissipation because of the negative extinction coefficient $\kappa$ (middle panel in Fig. 4(b)). An even more interesting situation occurs when a metallic screening ($\epsilon_1 < 0$) by photoexcited carriers coexists with an optical gain ($\epsilon_2 < 0$), where $\epsilon_1$ and $\epsilon_2$ stand for the real and imaginary parts of the dielectric constant, respectively. The refractive index $n = \left[\left(\sqrt{\epsilon_1^2 + \epsilon_2^2} + \epsilon_1\right)/2\right]^{1/2}$ then vanishes at the boundary between absorption and gain ($\epsilon_2 = 0$), which may further enhance the rapid spectral variation in $n$ (top panel in Fig. 4(c)). The group index correspondingly exceeds 300 at $\Delta t = -0.24$ ps (bottom panel in Fig. 4(d)), where a metallic screening ($\epsilon_1 < 0$) develops with the help of the SRLS itself. Remarkably, the extinction coefficient $\kappa = (\text{sgn}\,\epsilon_2)\left[\left(\sqrt{\epsilon_1^2 + \epsilon_2^2} - \epsilon_1\right)/2\right]^{1/2}$ remains negative in this gain region (middle panel in Fig. 4(c)), so that a probe wave does not decay in spite of the metallic character in $\epsilon_1$. An electromagnetic pulse therefore might be slowed down more than 300 times without loss under the present experimental condition.

Most previous studies of slow light generation used electromagnetically induced transparency [44-46,49] and photonic-band engineering [50] as the origin of a refractive index change $\Delta n$, which typically amounts to ~0.01 and ~0.1, respectively. In our case, by contrast, $\Delta n > 1$ is so large that $n$ even vanishes. A relatively large bandwidth $\Delta f \sim 0.5$ THz of the dispersion limits the achievable group refractive index here. This is not necessarily a disadvantage, because a broader dispersion allows a shorter pulse to be slowed down. In fact, photonic-band engineering emerged as a way to generate slow light with a broad bandwidth (~THz) [50], compared to a much narrower one (~kHz) achieved by electromagnetically induced transparency. Our experimental results show that lossless and broadband slow light generation is possible by simply shedding infrared light to a semimetal at room temperature. To avoid complication by transient effects, such as the temporal change from Fig. 4(b) to (c), continuous-wave or nanosecond $CO_2$ lasers promise better choice as the pump light source, though optical heating should be suppressed by efficient cooling. We expect SRLS to be robust against excitation-induced dephasing and scattering even for such a long-lasting driving, because its coherence time is determined by the relatively long carrier lifetime $T_1 = 8$ ps. The available bandwidth then becomes $\Delta f \sim 1/T_1 = 0.13$ THz, still keeping a relatively large value. We leave the implementation of slow light generation with this mechanism as a topic of future studies.

In summary, we performed ultrafast pump-probe spectroscopy on a $Cd_3As_2$ thin film in the multiterahertz frequency region, to find SRLS to dominate the transient absorption spectrum in the pump-probe overlap. Macroscopically, it originates from a transient grating with a blueshifted plasma frequency in the interfering pump and probe fields. The characteristic dispersive lineshape can be further traced back to microscopic optical transitions between the light-dressed electronic bands, the Floquet subbands, assisted by a diamagnetic coupling with the optical field. The concomitant sharp dispersion in the transient refractive index may be applicable to semimetal-based, lossless, broadband slow light generation at room temperature. These findings reveal a general aspect of light-matter interaction and lay the foundation of Floquet engineering for optical response of continuous energy bands. The application of circularly polarized driving fields promises an interesting future direction because of its ability to manipulate band topology and magnetic symmetry [8-10,51,52].

**Acknowledgements**

This work was supported by JST PRESTO (Grant Nos. JPMJPR20LA and JPMJPR2006), JST CREST (Grant No. JPMJCR20R4), and in part by JSPS KAKENHI (Grants Nos. JP19H01817 and JP20J01422, JP20H00343, and JP21K13852). Work at UCSB was supported by CATS Energy Frontier Research Center, which is funded by the U.S. Department of Energy, Basic Energy Sciences, under contract DE-AC02-06CH11357. R.M. also acknowledges partial support by Attosecond lasers for next frontiers in science and technology (ATTO) in Quantum Leap Flagship Program (MEXT Q-LEAP). A part of the computations and the FTIR measurement were performed using the facilities of the Supercomputer Center and the Materials Design and Characterization Laboratory, respectively, in The Institute for Solid State Physics, The University of Tokyo.

R.M. conceived this project. M.G. and S.S. fabricated the sample. N.K. and T.M. evaluated the linear response function. Y.M. and N.K. developed the pump-probe spectroscopy system with the help of J.Y., Y.K., and R.M. N.K. performed the experiment and analyzed the data. Y.M. conducted the phenomenological analysis. Y.M. and T.N.I. performed the microscopic calculations. All the authors discussed the results. Y.M. wrote the manuscript with the substantial help of N.K., T.N.I., and R.M., with the feedbacks from all the other coauthors.

**Figures and figure captions**

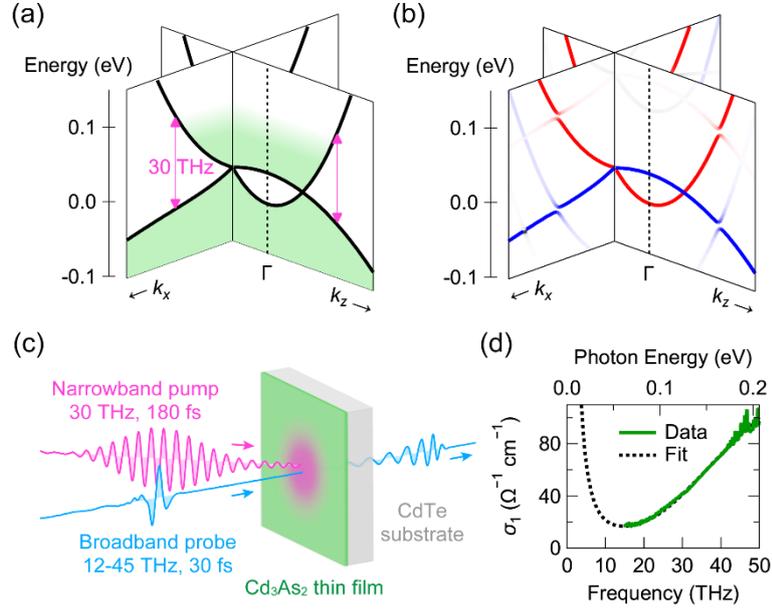

FIG. 1. (a) Band structure of $Cd_3As_2$ around the $\Gamma$ point [26]. (b) Schematic picture of the Floquet state formation by a periodic optical field. (c) Setup of the pump-probe experiment. (d) Optical conductivity of the sample. The model fitting (dotted line) takes into account the lower-frequency data outside the panel [25].

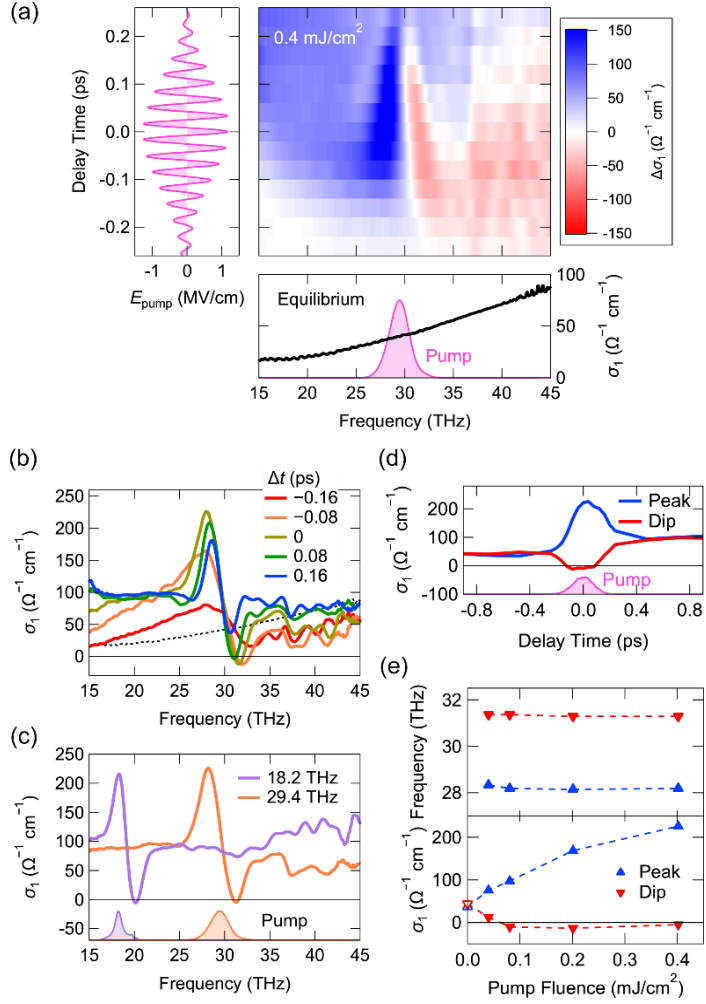

FIG. 2. (a) Change of the optical conductivity as a function of frequency (horizontal axis) and pump-probe delay time $\Delta t$ (vertical axis). Waveform of the pump pulse is shown on the left. The equilibrium optical conductivity is plotted on the bottom along with the pump power spectrum. Pump and probe pulses are collinearly polarized. (b) Transient optical conductivity at several delay times. The equilibrium spectrum is shown as a dotted line. (c), Optical conductivity at $\Delta t = 0.04$ ps for different pump frequencies, i.e., 29.4 THz (the same as in (a), (b)) and 18.2 THz (with a fluence of 0.25 mJ/cm$^2$, a peak electric field of 0.9 MV/cm). (d) Delay time dependence of the peak and dip values extracted from (a). Temporal profile of the pump intensity is shown as the shaded curve. (e) Top: Positions of the peak and the dip in optical conductivity at $\Delta t = 0.04$ ps, as a function of pump fluence. Bottom: A similar plot for the conductivity values at the peak and the dip. Equilibrium values at 28.2 and 31.3 THz are added as open triangles.

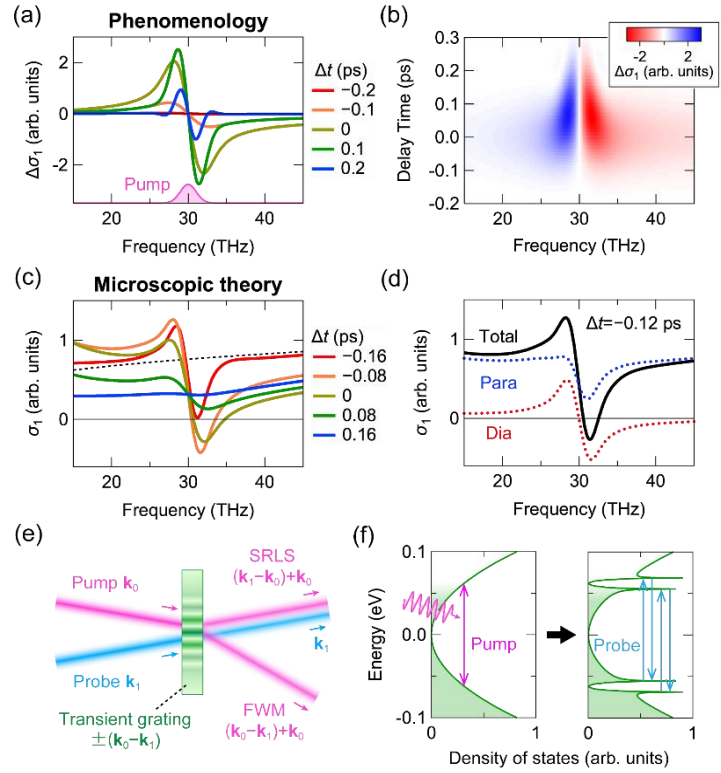

FIG. 3. (a) Change of the optical conductivity $\Delta\sigma_1$ calculated by a phenomenological model. Theoretical details are given in Supplemental Material [27]. (b) Two-dimensional plot of $\Delta\sigma_1$ as a function of frequency (horizontal axis) and pump-probe delay time (vertical axis). (c) Transient optical conductivity calculated by a microscopic model. (d) Contributions from the paramagnetic (blue) and diamagnetic (red) currents in the total optical conductivity (black) at $\Delta t = -0.12$ ps. (e) Geometric picture of stimulated Rayleigh scattering (SRLS) and four-wave mixing (FWM). $\mathbf{k}_0$ and $\mathbf{k}_1$ denote wavevectors of the pump and the probe, respectively. (f) SRLS induced by Floquet states in continuous bands. Ordinary ac Stark effect corresponds to transitions between the topmost and bottom peaks, and between the intermediate peaks, in the right panel.

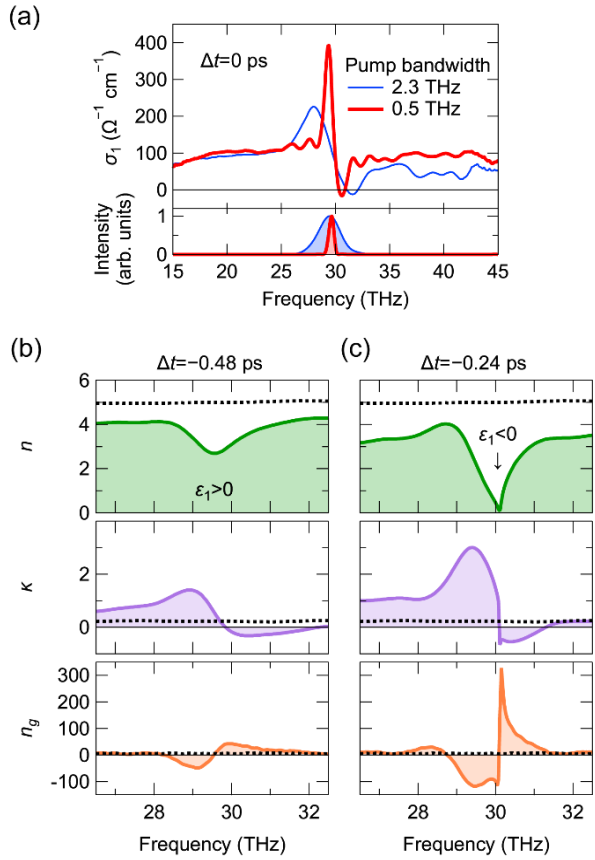

FIG. 4. (a) Transient optical conductivity for broader (thin) and narrower (thick) pump pulses. Pump power spectra are plotted on the bottom with their FWHM indicated in the figure. The broader pump is the same as in Fig. 2(a), while the narrower one has a pulse width of 0.88 ps, a fluence of 1.7 mJ/cm$^2$, and a peak electric field of 1.2 MV/cm. (b) Refractive index $n$ (top), extinction coefficient $\kappa$ (middle), and group refractive index $n_g$ (bottom) measured at $\Delta t = -0.48$ ps for the narrower pump in (a). Equilibrium spectra are shown as dotted lines. (c) The same data set for $\Delta t = -0.24$ ps.

# Supplemental Material for

# Stimulated Rayleigh Scattering Enhanced by a Longitudinal Plasma Mode in a Periodically Driven Dirac Semimetal $Cd_3As_2$

Yuta Murotani, Natsuki Kanda, Tatsuhiko N. Ikeda, Takuya Matsuda, Manik Goyal, Jun Yoshinobu, Yohei Kobayashi, Susanne Stemmer, and Ryusuke Matsunaga

## S1. Sample Preparation

A (112)-oriented $Cd_3As_2$ film was grown by molecular beam epitaxy on a (111)B CdTe substrate with a 3° miscut, as detailed elsewhere [S1]. The film thickness was 140-150 nm as determined by the growth time. At room temperature, the Hall mobility of the film was ~10,000 $cm^2$/Vs. The linear optical conductivity was measured by terahertz time-domain spectroscopy and Fourier-transform infrared spectroscopy and was fitted by a model function [S2].

## S2. Pump-probe spectroscopy

We used a Yb:KGW-based regenerative amplifier (center frequency 1030 nm, repetition rate 3 kHz, pulse energy 2 mJ, and pulse width 255 fs) as a light source. 80% of the output (1.6 mJ) was used to pump and seed two optical parametric amplifiers (OPAs) emitting signal beams with different colors (1350-1650 nm). Narrowband multiterahertz pump pulses were subsequently generated in a 500 μm-thick GaSe crystal through difference frequency generation [S3]. Inserting grating pair stretchers after the OPAs enabled us to vary the pump pulse width [S4] in the range of 180-880 fs, keeping a transform-limited waveform and a large pulse energy of ~1 μJ. The remaining part of the light source (0.4 mJ) was compressed to 14 fs with a throughput of 70 μJ, by a two-stage multi-plate broadening scheme [S5,S6]. The compressed pulses were further divided into two beams. One irradiated a 30 μm-thick GaSe crystal to generate broadband multiterahertz probe pulses through intra-pulse difference frequency generation. Their waveforms after transmitting the sample were measured by electro-optic sampling in a few-μm-thick GaSe flake on a diamond substrate, using the other beam as gate pulses [S2]. Typical waveforms and power spectra are shown in Fig. S1. The pump and probe beams were focused on the sample with spot sizes of >125 and 46 μm, respectively, with a relative angle of incidence, 20°. Penetration depth at the pump frequencies exceeded 3 μm, so that the film was uniformly excited in the probed region. In pump-probe measurements, the time difference between the pump and gate pulses was fixed and defined as the pump-probe delay time $\Delta t$, while the arrival time of the probe pulse was scanned to obtain electric field waveforms and transient response functions [S7]. We confirmed that another analysis method, i.e., fixing the time difference between the pump and the probe, and scanning the arrival time of the gate, returned qualitatively the same results including appearance of optical gain (see Section S5).

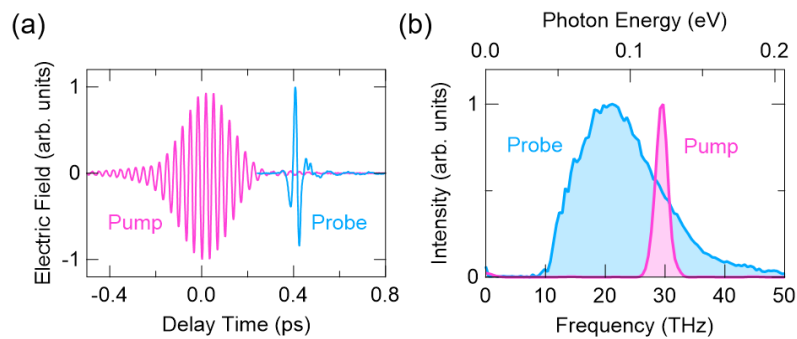

FIG. S1. (a) Waveforms and (b) power spectra of a typical pump pulse (tuned to 29.4 THz) and the probe pulse.

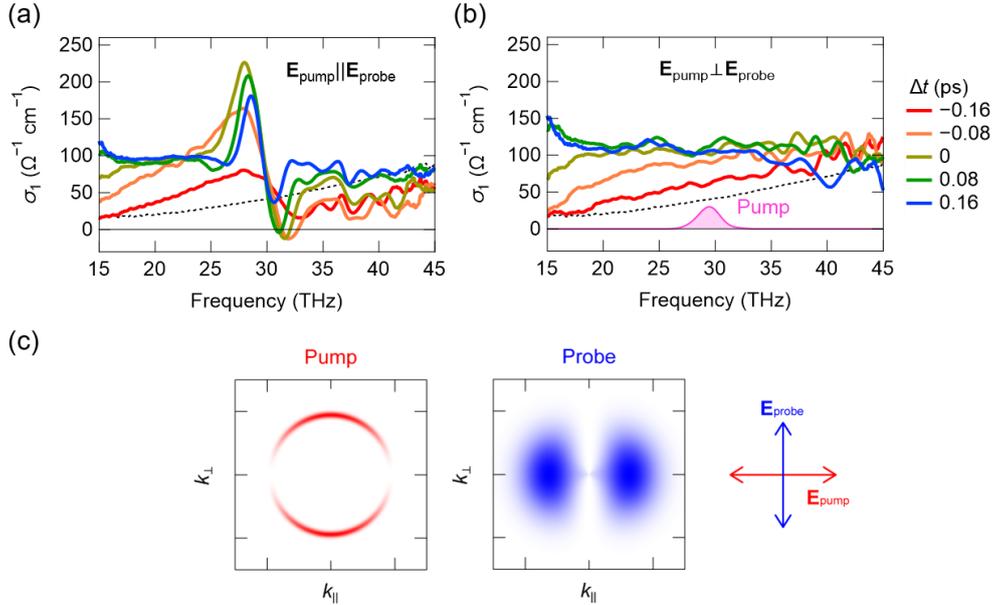

FIG. S2. (a), (b) Transient optical conductivity measured by probe pulses with a polarization direction parallel and perpendicular to the pump, respectively. Data in (a) is the same as that in Fig. 2(b) in the main text. The dotted lines are the equilibrium spectrum. (c), Distribution of carriers excited by the narrowband pump (left) and by the broadband probe (right). An ideal massless Dirac dispersion relation is assumed for simplicity, with the Dirac node located at the center. Polarization direction of each pulse is indicated by arrows.

**S3. Probe polarization dependence**

Figure S2 compares the transient optical conductivity measured by probe pulses with a polarization direction (a) parallel and (b) perpendicular to the pump. In stark contrast to S2(a) featured by a dispersive structure, only broad induced absorption appears in S2(b), with a spectral weight gradually shifting to the low-frequency side to develop a Drude-like response. Similar spectra were also observed in the case of near-infrared pump [S2]. This behavior is attributed to the real excitation of carriers which cause induced intraband and/or interband absorption. Absence of the dispersive structure is consistent with the interpretation of stimulated Rayleigh scattering (SRLS), since $\mathbf{E}_{\text{pump}}$ and $\mathbf{E}_{\text{probe}}$ perpendicular to each other drive interband transitions at different regions in the momentum space, as schematically shown in Fig. S2(c). Here, distributions of carriers excited by the pump (left) and the probe (right) are calculated for an ideal Dirac dispersion relation, neglecting all scattering processes. A horizontally polarized pulse (the pump in the figure) creates carrier distribution weighted vertically around Dirac nodes, and vice versa (for the probe).

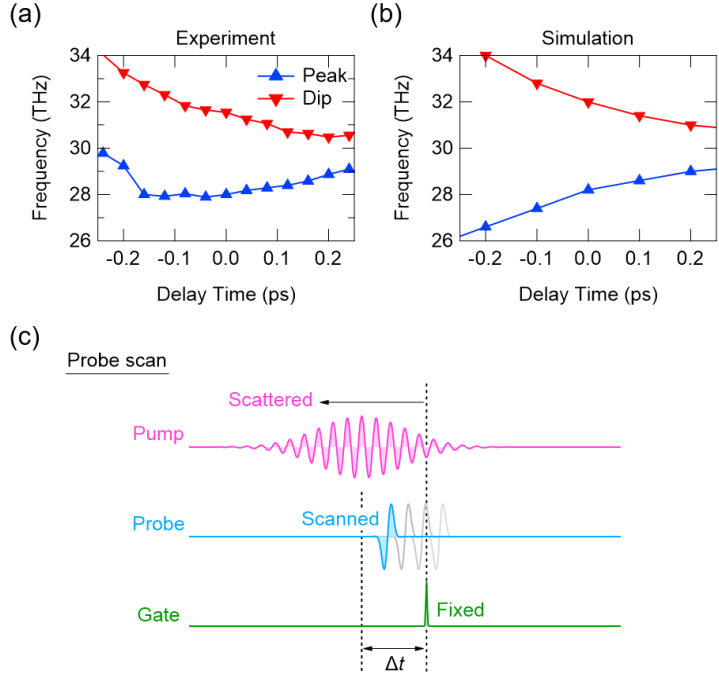

FIG. S3. (a) Frequencies of the peak and the dip extracted from the experimental data in Fig. 2(a). (b) The corresponding simulation result extracted from Fig. 3(b). (c), Schematic of the probe scan.

## S4. Temporal change in the peak-to-dip separation

Figures 2(a) and 2(b) in the main text exhibit a narrowing down of the separation between the peak and the dip for increasing pump-probe delay time. This behavior is more clearly seen in Fig. S3(a), which plots the temporal variation of the frequencies at the peak and the dip. As already discussed in the main text, the macroscopic model of SRLS reasonably reproduces the experimental result (Fig. S3(b)).

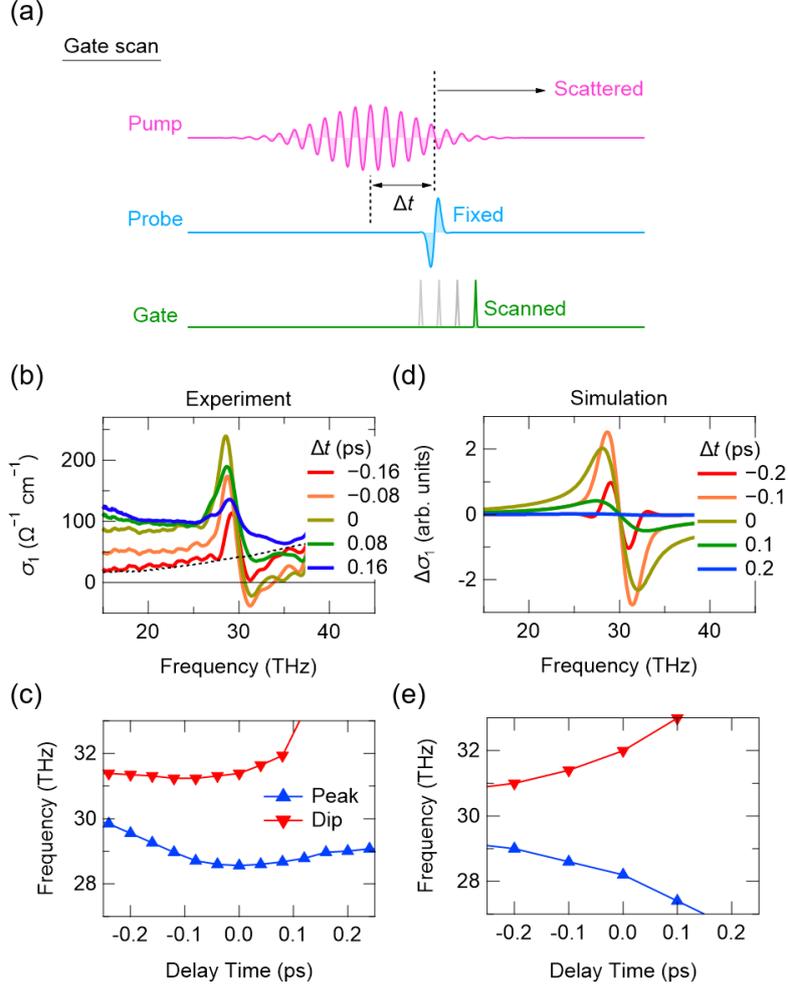

FIG. S4. (a) Schematic of the gate scan. (b) Transient optical conductivity obtained by the gate scan. The excitation condition is the same as in Figs. 2(a)-(c) in the main text. (c) Frequencies of the peak and the dip extracted from the data in (b). (d) Change in the optical conductivity simulated by the macroscopic model. (e) Peak and dip frequencies extracted from the simulation result in (d).

This narrowing down of the peak-to-dip separation in time is attributed to the measurement scheme of "probe scan" in our experiment. As sketched in Fig. S3(c), the time difference between the pump and gate pulses is fixed and defined as the "pump-probe" delay time $\Delta t$, and the arrival time of the probe pulse is scanned to obtain a waveform. In this configuration, causality allows only a part of the pump pulse before the arrival of the gate (indicated by a one-way arrow in Fig. S3(c)) to be scattered and detected. As a result, an effective bandwidth of the pump pulse decreases as $\Delta t$ increases, leading to a narrowing down of the dispersive lineshape.

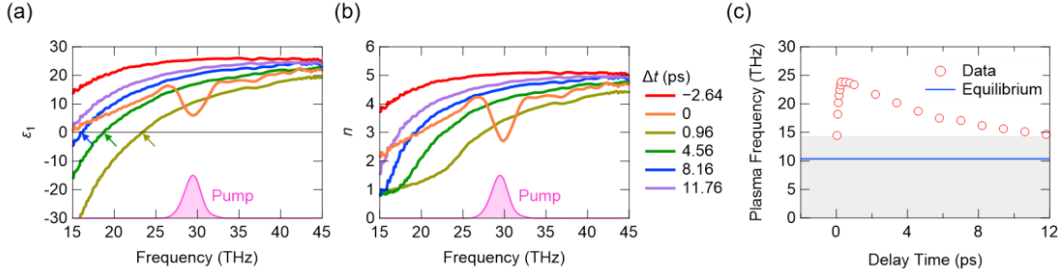

FIG. S5. (a) Time-resolved dielectric function and (b) refractive index. Position of the screened plasma frequency is indicated by arrows in (a). Experimental conditions are the same as those in Fig. 2(b) in the main text. (c) Temporal change of the screened plasma frequency, extracted from the data in (a). The equilibrium value is shown by the solid line. The shaded region is outside the detected range.

As seen in Fig. 2(b) and in Fig. 3(a), the peak and the dip not only get closer but sharpen for increasing delay time. The mechanism discussed above also accounts for this behavior.

## S5. Probe scan vs. gate scan

One can also adopt another measurement scheme: time difference between the pump and the probe is fixed and the arrival time of the gate is scanned (Fig. S4(a)). This "gate scan" procedure is a natural extension of conventional pump-probe spectroscopy with a spectrometer. Both methods have their own merits and demerits. On one hand, the probe scan has an advantage in time resolution, because the duration of the ultrashort gate pulse determines it [S7]. However, this procedure may result in unphysical spectra around the pump-probe overlap [S8]. On the other hand, the gate scan is not good at time resolution, because the relatively long probe THz pulse determines it. However, this procedure directly measures the absorption and refraction experienced by a probe pulse, so that interpretation is simpler. We have compared these two methods in our experiment. In the latter configuration, we observe again a dispersive lineshape reaching an optical gain at the largest change (Fig. S4(b)). Here, however, the separation between the peak and the dip tends to widen as $\Delta t$ increases (Fig. S4(c)), unlike the probe scan. This behavior is indeed natural, because only a part of the pump pulse after irradiation by the probe (indicated by a one-way arrow in Fig. S4(a)) can be scattered by the transient grating. The macroscopic model of SRLS again reproduces this trend when the gate scan is simulated (Figs. S4(d) and S4(e)).

## S6. Light-induced shift of the plasma frequency

Figure S5(a) shows the time-resolved dielectric function $\epsilon_1$ for the same data set as in Fig. 2(b) in the main text. The zero in $\epsilon_1$, aside from those by phonons, gives the screened plasma frequency. In equilibrium ($\Delta t = -2.64$ ps), it is located at 10 THz in our sample [25], which is outside the range detected here. During the photoexcitation ($\Delta t = 0$ ps), a clear dip at the pump frequency emerges, which is a characteristic of SRLS. Meanwhile the screened plasma frequency enters the plotted region, and gradually turns back to the lower frequency side after photoexcitation ($\Delta t > 0.96$ ps). The shift of the plasma frequency is accompanied with an overall reduction of the refractive index, as shown in Fig. S5(b). Figure S5(c) follows the temporal change of the screened plasma frequency, which is proportional to the square root of the carrier density. The squared plasma frequency decays with a time constant of 7 ps, which is consistent with the decay of photoexcited carriers in Ref. [25].

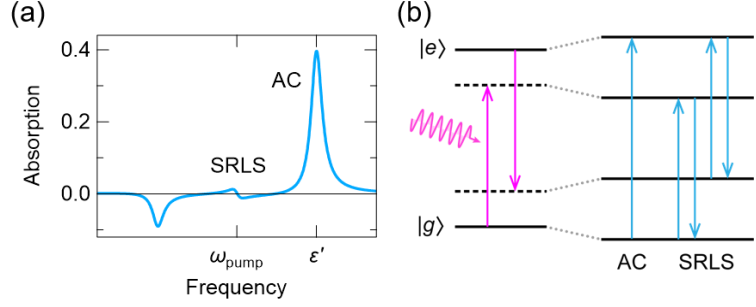

FIG. S6. (a) Typical absorption spectrum of a two-level system irradiated by a monochromatic light wave [S9]. AC and SRLS denote ac Stark effect and stimulated Rayleigh scattering, respectively. (b) Level diagram of the dressed states and the resulting optical transitions. $|g\rangle$ and $|e\rangle$ designate the ground and excited states, respectively.

### S7. Stimulated Rayleigh scattering in two-level systems

Figure S6(a) shows a typical absorption spectrum of a two-level system irradiated by a monochromatic light wave [S9]. SRLS accompanies the well-known ac Stark effect (AC). Optical gain appearing below the pump frequency ($\omega_{\text{pump}}$) is the three-photon resonance, arising from a higher-order nonlinearity. Energy levels of the original and dressed states, and the related optical transitions, are shown in Fig. S6(b).

### S8. Phenomenological model

For the phenomenological description of SRLS, we used a two-beam coupling framework [S9]. Suppose that local optical intensity $I(\mathbf{r}, t)$ proportionally induces the refractive index change $n_{\text{NL}}(\mathbf{r}, t)$, which decays with a time constant of $\tau$. Spatiotemporal variation of $n_{\text{NL}}$ is then determined by

$$\tau \frac{\partial n_{\text{NL}}}{\partial t} + n_{\text{NL}} = n_2 I, \qquad (1)$$

where $n_2$ is a material-dependent constant. As discussed in the main text, $n_2$ is negative in the present case because photoexcited carriers reduce the refractive index around 30 THz with a lifetime of $\tau = 8$ ps due to a blueshift of the plasma frequency from 10 THz [S2]. The nonlinear polarization is given by

$$\mathbf{P}^{\mathrm{NL}} = 2n_0\epsilon_0 n_{\mathrm{NL}}\mathbf{E}, \tag{2}$$

where $n_0$ is the refractive index at equilibrium, $\epsilon_0$ the vacuum permittivity, and $\mathbf{E}$ the electric field in the film. Now the pump and probe fields are introduced via

$$\mathbf{E}(\mathbf{r},t) = \mathbf{E}_0(t)e^{i\mathbf{k}_0\cdot\mathbf{r}} + \mathbf{E}_1(t)e^{i\mathbf{k}_1\cdot\mathbf{r}} + \mathrm{c.c.}, \tag{3}$$

where $\mathbf{E}_{0(1)}(t)$ denotes the rotating wave component of the pump (probe) waveform, $\mathbf{k}_{0(1)}$ the wave vector, and c.c. the complex conjugate. Optical intensity is then given by

$$\begin{aligned}I(\mathbf{r},t) &= n_0\epsilon_0 c\overline{\mathbf{E}(\mathbf{r},t)^2}\\ &= 2n_0\epsilon_0 c\{|\mathbf{E}_0(t)|^2 + |\mathbf{E}_1(t)|^2 + [\mathbf{E}_1(t)\cdot\mathbf{E}_0^*(t)e^{i(\mathbf{k}_1-\mathbf{k}_0)\cdot\mathbf{r}} + \mathrm{c.c.}]\},\end{aligned} \tag{4}$$

where $c$ indicates the speed of light and the overline a time average over several periods. The resulting nonlinear polarization contains a component that propagates in the same direction as the probe field,

$$\mathbf{P}_1^{\mathrm{NL}}(t) = 2n_0\epsilon_0[n_{\mathrm{NL}}^{00}(t)\mathbf{E}_1(t) + n_{\mathrm{NL}}^{10}(t)\mathbf{E}_0(t)], \tag{5}$$

where $n_{\mathrm{NL}}^{00}$ and $n_{\mathrm{NL}}^{10}$ obey

$$\begin{aligned}\tau\frac{dn_{\mathrm{NL}}^{00}}{dt} + n_{\mathrm{NL}}^{00} &= 2n_0 n_2\epsilon_0 c|\mathbf{E}_0(t)|^2,\\ \tau\frac{dn_{\mathrm{NL}}^{10}}{dt} + n_{\mathrm{NL}}^{10} &= 2n_0 n_2\epsilon_0 c\mathbf{E}_1(t)\cdot\mathbf{E}_0^*(t).\end{aligned} \tag{6}$$

Among these, $n_{\mathrm{NL}}^{10}$ is induced by spatial interference of the pump and probe fields and diffracts the pump beam into the propagation direction of the probe, which is nothing but SRLS. The resulting change in the effective susceptibility $\Delta\chi$ experienced by the probe pulse is obtained by Fourier transforming $\mathbf{P}_1^{\mathrm{NL}}(t)$ against the probe delay time and dividing it by the Fourier component of $\mathbf{E}_1(t)$. Change in the optical conductivity is calculated from the relation $\Delta\sigma_1 = \omega\epsilon_0\,\mathrm{Im}\,\Delta\chi$. Figures 3(a) and 3(b) show the simulated results. This effect disappears for orthogonally polarized pump and probe fields due to an implicit assumption of optical isotropy, which is reasonable for the pseudo-hexagonal (112) plane of $Cd_3As_2$.

## S9. Microscopic theory

For microscopic analysis, we adopted an effective two-band model applicable to the low-energy band structure. The Hamiltonian reads

$$H_{\mathrm{eff}}(\mathbf{k}) = \epsilon_0(\mathbf{k}) + \begin{pmatrix} M(\mathbf{k}) & D(\mathbf{k}) & 0 & 0 \\ D^*(\mathbf{k}) & -M(\mathbf{k}) & 0 & 0 \\ 0 & 0 & M(\mathbf{k}) & -D^*(\mathbf{k}) \\ 0 & 0 & -D(\mathbf{k}) & -M(\mathbf{k}) \end{pmatrix}, \tag{7}$$

where $\epsilon_0(\mathbf{k}) = C_0 + C_1 k_z^2 + C_2(k_x^2 + k_y^2) + C_3 k_z^4 + C_4(k_x^2 + k_y^2)^2 + C_5(k_x^2 + k_y^2)k_z^2 + C_6 k_x^2 k_y^2$, $M(\mathbf{k}) = M_0 + M_1 k_z^2 + M_2(k_x^2 + k_y^2) + M_3 k_z^4 + M_4(k_x^2 + k_y^2)k_z^2 + M_5(k_x^2 + k_y^2)^2 + M_6 k_x^2 k_y^2$, and $D(\mathbf{k}) = (D_0 + D_1 k_z^2 + D_2 k_x^2 + D_3 k_y^2)k_x - i(D_0 + D_1 k_z^2 + D_2 k_y^2 + D_3 k_x^2)k_y$.

Here, we have extended the second-order model in the literature [S10] to the fourth order in $\mathbf{k}$, because the second-order model underestimates the interband transition dipole moment at multiterahertz frequencies. We fitted the dispersion relation predicted by a second-order eight-band Kane model [S10] by the above model, to obtain $C_0 = -0.0107$ eV, $C_1 = 11.84$ eV Å$^2$, $C_2 = 17.66$ eV Å$^2$, $C_3 = 592.7$ eV Å$^4$, $C_4 = -413.9$ eV Å$^4$, $C_5 = -640.5$ eV Å$^4$, $C_6 = -680.1$ eV Å$^4$, $M_0 = -0.0215$ eV, $M_1 = 19.80$ eV Å$^2$, $M_2 = 21.59$ eV Å$^2$, $M_3 = 592.7$ eV Å$^4$, $M_4 = -672.0$ eV Å$^4$, $M_5 = 358.2$ eV Å$^4$, $M_6 = -203.9$ eV Å$^4$, $D_0 = 1.137$ eV Å, $D_1 = -133.8$ eV Å$^3$, $D_2 = 69.3$ eV Å$^3$, and $D_3 = -125.7$ eV Å$^3$. The optical field is minimally coupled to the system through Peierls substitution $\mathbf{k} \to \mathbf{k} + e\mathbf{A}/\hbar$, where $e(>0)$ denotes the elementary charge and $\mathbf{A}$ the vector potential. To minimize computational efforts, we restricted ourselves to the optical field polarized in the $z$ direction, i.e., $\mathbf{A} = (0,0,A)$. In addition, we assumed the system to be isotropic in the $xy$ plane, by replacing $C_6 = 2C_4$, $M_6 = 2M_4$, and $D_3 = D_2$. We believe that this simplification does not matter significantly, because Cd$_3$As$_2$ is almost optically isotropic [S11,S12]. After diagonalization, one obtains an effective $2 \times 2$ Hamiltonian

$$H(\mathbf{k}) = \mathcal{E}(\mathbf{k}) - J^z(\mathbf{k})A + \frac{K^{zz}(\mathbf{k})}{2}A^2 + \cdots, \tag{8}$$

with the two-fold degeneracy. The field-free part is given by

$$\mathcal{E}(\mathbf{k}) = \begin{pmatrix} \epsilon_1(\mathbf{k}) & 0 \\ 0 & \epsilon_2(\mathbf{k}) \end{pmatrix}, \tag{9}$$

where $\epsilon_{1,2}(\mathbf{k}) = \epsilon_0(\mathbf{k}) \pm \sqrt{M(\mathbf{k})^2 + |D(\mathbf{k})|^2}$. The paramagnetic current operator $J^z(\mathbf{k})$ is responsible for optical transitions in the linear response. The second-order coupling constant $K^{zz}(\mathbf{k})$ arises from the parabolicity or the inverse effective mass of energy bands; more specifically, $\mathbf{k} \cdot \mathbf{p}$ perturbation theory tells us that

$$K^{zz}_{11(22)}(\mathbf{k}) = \frac{e^2}{\hbar^2}\frac{\partial^2 \epsilon_{1(2)}(\mathbf{k})}{\partial k_z^2} - \frac{2|J^z_{12}(\mathbf{k})|^2}{\epsilon_{1(2)}(\mathbf{k}) - \epsilon_{2(1)}(\mathbf{k})}, \tag{10}$$

where the first term on the right-hand side is proportional to the inverse effective mass. This second-order coupling gives rise to the diamagnetic contribution to the total current operator,

$$J^z_{\text{tot}}(\mathbf{k}) = -\frac{\partial H(\mathbf{k})}{\partial A} = J^z(\mathbf{k}) - K^{zz}(\mathbf{k})A + \cdots. \tag{11}$$

Upon this two-band basis, we let the electron population $n_{1,2}(\mathbf{k})$ and the interband polarization $P(\mathbf{k})$ follow

$$\frac{\partial n_1(\mathbf{k})}{\partial t} = -\frac{2}{\hbar}\text{Im}[H_{21}(\mathbf{k})P(\mathbf{k})] - \frac{n_1(\mathbf{k}) - f_1(\mathbf{k})}{T_1'} - \frac{n_1(\mathbf{k}) - f_1^{\text{eq}}(\mathbf{k})}{T_1}, \quad (12)$$

$$\frac{\partial n_2(\mathbf{k})}{\partial t} = +\frac{2}{\hbar}\text{Im}[H_{21}(\mathbf{k})P(\mathbf{k})] - \frac{n_2(\mathbf{k}) - f_2(\mathbf{k})}{T_1'} - \frac{n_2(\mathbf{k}) - f_2^{\text{eq}}(\mathbf{k})}{T_1}, \quad (13)$$

$$i\hbar\frac{\partial P(\mathbf{k})}{\partial t} = [H_{11}(\mathbf{k}) - H_{22}(\mathbf{k})]P(\mathbf{k}) - H_{12}(\mathbf{k})[n_1(\mathbf{k}) - n_2(\mathbf{k})] - i\hbar\frac{P(\mathbf{k})}{T_2}. \quad (14)$$

The terms that contain $T_1$, $T_1'$, and $T_2$ describe scattering effects within a relaxation time approximation. Electron-electron scattering redistributes an initially nonthermal electron population into a time-dependent Fermi distribution function $f_{1,2}(\mathbf{k})$, within a time interval of $T_1'$. The targeted chemical potential and electron temperature are determined from the internal energy of electrons,

$$U = \sum_{\mathbf{k}} [\epsilon_1(\mathbf{k})n_1(\mathbf{k}) + \epsilon_2(\mathbf{k})n_2(\mathbf{k})], \quad (15)$$

where the degeneracy factor of 2 is omitted for simplicity. Electron-phonon scattering subsequently equilibrates the electron and lattice subsystems within a time of $T_1$, after which electrons settle in a static distribution function $f_{1,2}^{\text{eq}}(\mathbf{k})$. Scattering events also destroy the interband polarization with a time constant of $T_2$. We note that the above modeling of scattering terms breaks the gauge invariance. For example, adding a constant term in **A** may change the final results. Gauge-invariant formulation of the relaxation terms is possible, e.g., on the basis of instantaneous eigenstates [S13], though we do not aim at such elaboration. We chose $T_1' = 500$ fs and $T_2 = 20$ fs, and neglected the $T_1$ relaxation terms because $T_1 = 8$ ps far exceeds the time scale of interest. Current density can be expressed as a sum of paramagnetic, diamagnetic, and higher-order terms,

$$j^z = j_{\text{para}}^z + j_{\text{dia}}^z + \cdots, \quad (16)$$

where

$$j_{\text{para}}^z = \sum_{\mathbf{k}} [J_{21}^z(\mathbf{k})P(\mathbf{k}) + J_{12}^z(\mathbf{k})P^*(\mathbf{k}) + J_{11}^z(\mathbf{k})n_1(\mathbf{k}) + J_{22}^z(\mathbf{k})n_2(\mathbf{k})],$$

$$j_{\text{dia}}^z = -\sum_{\mathbf{k}} [K_{21}^z(\mathbf{k})P(\mathbf{k}) + K_{12}^z(\mathbf{k})P^*(\mathbf{k}) + K_{11}^{zz}(\mathbf{k})n_1(\mathbf{k}) + K_{22}^{zz}(\mathbf{k})n_2(\mathbf{k})]A. \quad (17)$$

Pump and probe fields were introduced through $A = A_0 + A_1$, where the subscripts 0 and 1 denote pump and probe, respectively. To remove the contribution from four-wave mixing, we averaged the probe-induced current density over the phase of the pump waveform [S14]. The resulting current density was Fourier-transformed with respect to the real time $t$, and divided by the Fourier transform of the probe electric field $E_1 = -\partial A_1/\partial t$, to obtain transient optical conductivity. This procedure differs from the experimental setup in the main text, where the probe delay time (not the real time) was scanned to perform Fourier transform. However, as shown in Section S5, Fourier transform with respect to the gate delay time in the experiment (corresponding to the real time in theory) led to essentially the same results. Therefore, we concentrated on Fourier transform with respect to the real time, which is numerically more feasible. The vector potentials were taken as

$$A_i(t) = a_i \cos(2\pi f_i t + \phi_i) \exp\left[-2\ln 2 \left(\frac{t-t_i}{w_i}\right)^2\right] \quad (i = 0,1), \tag{18}$$

where $A_0(t)$ and $A_1(t)$ denote the pump and probe fields, respectively. Considering a typical experimental condition, we chose the parameters as $ea_0/\hbar = 0.24\,\text{nm}^{-1}$, $f_0 = 30\,\text{THz}$, and $w_0 = 180\,\text{fs}$ for the pump and $ea_1/\hbar = 0.05\,\text{nm}^{-1}$, $f_1 = 20\,\text{THz}$, and $w_1 = 20\,\text{fs}$ for the probe. We fixed $t_0 = 0$ and varied the probe delay $t_1 = \Delta t$. To remove the contribution from the four-wave mixing, we averaged the field-induced electric current over the relative phase $\phi_0 - \phi_1$. Specifically, we fixed $\phi_1 = 0$ and averaged the results over $\phi_0 = (2\pi/N_\phi)m$ ($m = 1,2,\ldots,N_\phi$) for $N_\phi = 8$. We confirmed that $N_\phi = 16$ little changed the results.

We numerically solved the time-dependent Schrödinger equation in the presence of the vector potentials. We considered a large $\mathbf{k}$-space sphere $|\mathbf{k}| \leq k_{\max}$, in which each $\mathbf{k}$-point is parametrized as $\mathbf{k} = k(\sin\theta\cos\varphi, \sin\theta\sin\varphi, \cos\theta)$ with $\theta$ and $\varphi$ being the polar and azimuthal angles, respectively. We chose $k_{\max} = 1\,\text{nm}^{-1}$ large enough that appreciable interband transitions occur only within the sphere. Since the cylindrical symmetry of the model guarantees the $\varphi$-independence of the response, we discretized the system into $100 \times 30$ points in the $k$-$\theta$ plane, for $0 \leq k \leq k_{\max}$ and $0 \leq \theta \leq \pi$ (we confirmed the convergence of the results). We solved the set of equations with the 4th-order Runge-Kutta algorithm and calculated the total electric current by summing contributions from all $\mathbf{k}$-points. Note that, in the $T_1'$ relaxation terms, $f_{1,2}(\mathbf{k})$ are determined by the time-dependent temperature $T(t)$ and chemical potential $\mu(t)$, which are common for both bands. During time evolution, $T(t)$ and $\mu(t)$ were solved self-consistently with the system's internal energy $U(t)$ and the total number of electrons that was kept constant throughout the simulation.

## S10. Derivation of the macroscopic model from the microscopic one

Figure S7 shows the outline of the derivation. We start from the microscopic Eqs. (12)-(14). Within the rotating-wave approximation, the electron-hole pair density $\Delta N = \sum_\mathbf{k}[n_1(\mathbf{k}) - f_1^{\text{eq}}(\mathbf{k})] = \sum_\mathbf{k}[f_2^{\text{eq}}(\mathbf{k}) - n_2(\mathbf{k})]$ follows

$$\frac{\partial \Delta N}{\partial t} = \frac{2}{\hbar} \text{Im}\left[A_-(t) \sum_\mathbf{k} J_{21}(\mathbf{k}) P(\mathbf{k})\right] - \frac{\Delta N}{T_1}, \tag{19}$$

where $A_\pm(t)$ denotes the rotating and counter-rotating components of the vector potential $A(t)$. We have neglected pair generation and recombination caused by the carrier-carrier scattering, or the $T_1'$ terms. In the lowest order, $P(\mathbf{k})$ is given by

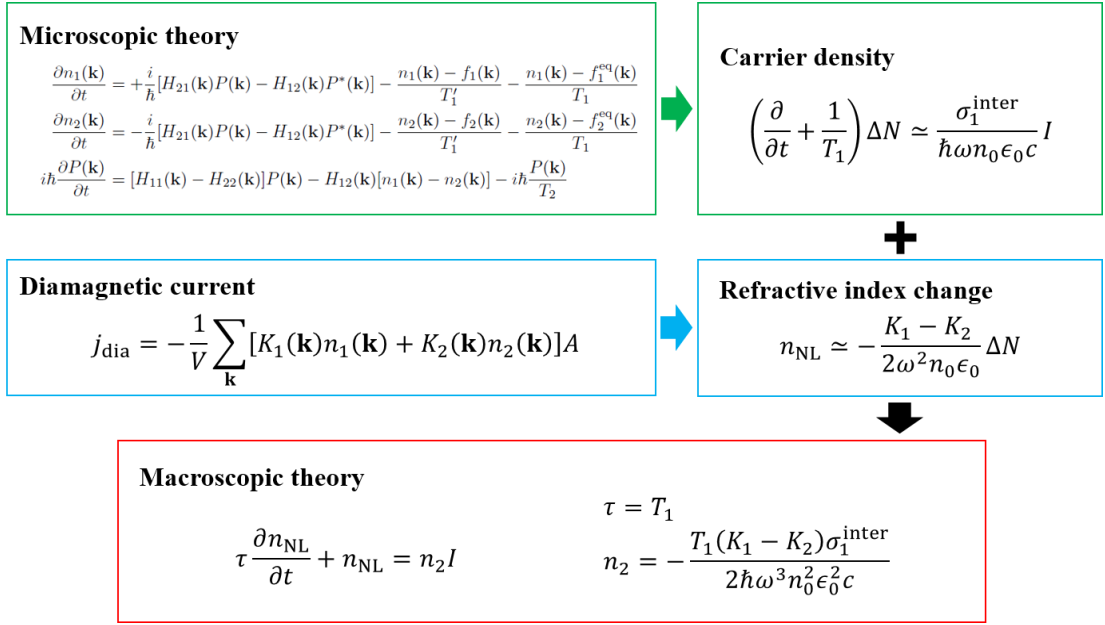

FIG. S7. Derivation of the macroscopic model from the microscopic one.

$$P(\mathbf{k}) = \int_0^\infty \frac{d\omega}{2\pi} A(\omega) e^{-i\omega t} \frac{[f_1^{eq}(\mathbf{k}) - f_2^{eq}(\mathbf{k})] J_{12}(\mathbf{k})}{\hbar(\omega + i/T_2) - \epsilon_1(\mathbf{k}) + \epsilon_2(\mathbf{k})}, \quad (20)$$

where $A(\omega)$ is the spectral amplitude of $A(t)$. Using Eq. (6), we obtain

$$\sum_{\mathbf{k}} J_{21}(\mathbf{k}) P(\mathbf{k}) = \int_0^\infty \frac{d\omega}{2\pi} A(\omega) e^{-i\omega t} \sum_{\mathbf{k}} \frac{[f_1^{eq}(\mathbf{k}) - f_2^{eq}(\mathbf{k})] |J_{12}(\mathbf{k})|^2}{\hbar(\omega + i/T_2) - \epsilon_1(\mathbf{k}) + \epsilon_2(\mathbf{k})}$$
$$\simeq A_+(t) \sum_{\mathbf{k}} \frac{[f_1^{eq}(\mathbf{k}) - f_2^{eq}(\mathbf{k})] |J_{12}(\mathbf{k})|^2}{\hbar(\omega_0 + i/T_2) - \epsilon_1(\mathbf{k}) + \epsilon_2(\mathbf{k})}, \quad (21)$$

where $\omega$ in the denominator is replaced by $\omega_0$, a representative frequency included in $A(t)$. Substitution of Eq. (21) into Eq. (19) yields

$$\frac{\partial \Delta N}{\partial t} = \frac{\omega_0 \sigma_1^{\text{inter}}(\omega_0)}{\hbar} \overline{A(t)^2} - \frac{\Delta N}{T_1}, \quad (22)$$

where the interband optical conductivity,

$$\sigma_1^{\text{inter}}(\omega) = \frac{1}{\omega} \text{Im} \sum_{\mathbf{k}} \frac{[f_1^{eq}(\mathbf{k}) - f_2^{eq}(\mathbf{k})] |J_{12}(\mathbf{k})|^2}{\hbar(\omega + i/T_2) - \epsilon_1(\mathbf{k}) + \epsilon_2(\mathbf{k})}, \quad (23)$$

has been defined. Using $\overline{A(t)^2} \simeq \overline{E(t)^2}/\omega_0^2$ and $I(t) = n_0 \epsilon_0 c \overline{E(t)^2}$, we obtain

$$\frac{\partial \Delta N}{\partial t} = \frac{\sigma_1^{\text{inter}}(\omega_0)}{\hbar \omega_0 n_0 \epsilon_0 c} I - \frac{\Delta N}{T_1}. \quad (24)$$

This is the well-known rate equation for electron-hole pair excitation.

We next turn to the diamagnetic current density,

$$j_{\text{dia}} = -\sum_{\mathbf{k}}[K_1(\mathbf{k})n_1(\mathbf{k}) + K_2(\mathbf{k})n_2(\mathbf{k})]A. \tag{25}$$

The interband matrix elements are omitted because numerical analysis shows their minor importance. Neglecting $\mathbf{k}$-dependence of $K_{1,2}(\mathbf{k})$, the light-induced change in $j_{\text{dia}}$ is given by

$$\Delta j_{\text{dia}} = -(K_1 - K_2)\Delta N \cdot A. \tag{26}$$

This relation is alternatively expressed in terms of the electric polarization,

$$\Delta P_{\text{dia}} = \int_{-\infty}^{t} dt' \, \Delta j_{\text{dia}}(t')$$

$$\simeq -\frac{1}{\omega_0^2}(K_1 - K_2)\Delta N \cdot E, \tag{27}$$

where the second equality derives from the slow variation in $\Delta N$ compared to the fast oscillation in $A$ or $E$. Equating the above expression with $\Delta P_{\text{dia}} = 2n_0\epsilon_0 n_{\text{NL}} E$, the refractive index change is given by

$$n_{\text{NL}} = -\frac{K_1 - K_2}{2\omega_0^2 n_0 \epsilon_0} \Delta N. \tag{28}$$

Now we are ready to end up with the macroscopic model, by combining Eqs. (24) and (28) to obtain

$$\tau \frac{\partial n_{\text{NL}}}{\partial t} + n_{\text{NL}} = n_2 I, \tag{29}$$

with

$$\tau = T_1, \tag{30}$$

$$n_2 = -\frac{T_1(K_1 - K_2)\sigma_1^{\text{inter}}(\omega_0)}{2\hbar\omega_0^3 n_0^2 \epsilon_0^2 c}. \tag{31}$$

A factor $\omega_0^3$ in the denominator of Eq. (31) suggests that lower frequency is preferred for SRLS, as long as the intraband response does not obscure it. This dependence stems partly from (24) and partly from (28). First, Eq. (24) shows that a larger number of carriers are generated by lower-frequency fields, because the optical energy is distributed to a larger number of photons to be absorbed. Second, Eq. (28) indicates that the injected carriers change the refractive index more efficiently at lower frequencies. This behavior is easily expected from the Drude model, which predicts a frequency-dependent change in the dielectric constant,

$$\Delta\epsilon = -\left(\frac{1}{m_e} + \frac{1}{m_h}\right)\frac{e^2\Delta N}{\epsilon_0 \omega^2}, \tag{32}$$

at frequencies higher than the damping rate. Here, $m_e$ and $m_h$ denote the effective mass of electrons and holes, respectively. $\mathbf{k} \cdot \mathbf{p}$ perturbation theory tells us that the inverse effective mass in the above equation is contributed by both the diamagnetic current and the paramagnetic current, the latter arising from virtual excitation of interband transitions. At high frequencies, the paramagnetic current fails to contribute to the inverse effective mass because interband transitions are no longer virtual. In such a situation, $e^2/m_e$ and $e^2/m_h$ in Eq. (32) are replaced by $K_1$ and $-K_2$, so that Eqs. (27) and (28) follow. We note that the frequency dependence of Eq. (31) is consistent with the experimental tendency seen in Fig. 2(c), where a weaker driving (0.9 MV/cm) at a lower frequency (18.2 THz) and a stronger driving (1.3 MV/cm) at a higher frequency (29.4 THz) led to absorption changes with a similar magnitude, though saturation effect hinders quantitative comparison.

The paramagnetic current density, neglected so far, has been extensively studied in the conventional nonlinear optics. Therefore, we only discuss its consequences without equations. It is known that the paramagnetic current also leads to SRLS in two-level systems [S9]. In this case, however, relative positions of the peak and the dip are inverted depending on the sign of detuning, so that the dispersive lineshape tends to be cancelled out when integrated over a continuum. Spectral hole burning, arising from a combination of Pauli blocking and ac Stark effect, survives such a cancellation. Our numerical analysis revealed the tendency of hole burning to be suppressed when the dephasing time $T_2$ is short. It is also possible that the diamagnetic coupling $K_{1,2}$ is so large that it can hide the spectral hole burning. Since either the value of $T_2$ or the correct parameters in the effective Hamiltonian are not known, these two possibilities are not resolved at present.

## S11. Achievable group delay

Efficiency of slow light generation is often assessed by the delay-bandwidth product, $\text{DBP} = \Delta t \cdot \Delta f \simeq (L/\lambda)\Delta n$, and its normalized form, $\Delta n$, where $\Delta t = n_g L/c \simeq L\Delta n/\lambda \Delta f$ is the optical delay given by a material with a length of $L$, $\Delta f$ the bandwidth in which $n_g$ is large and flat, $\lambda$ the wavelength, and $\Delta n$ the refractive index change [S15]. DBP approximately measures the optical delay ($\Delta t$) in units of the shortest pulse width available for slow light ($\sim 1/\Delta f$), so that a large value is preferred (typically several tens). For the same reason, large $\Delta n$ is desirable. Most previous studies used electromagnetically induced transparency [S16] and photonic band engineering [S15] as the origin of rapid spectral variations in $n$, leading typically to $\Delta n \sim 0.01$ and $\Delta n \sim 0.1$, respectively. In our case, by contrast, $\Delta n \sim 1$ is so large that $n$ even vanishes. Unfortunately, a small film thickness $L = 140$ nm and a relatively large wavelength $\lambda = 10$ μm limit DBP here, giving $\text{DBP} = 0.06$ for $\Delta n = 4$. However, it is possible to increase the propagation length $L$ to a few μm, because the penetration depth at this wavelength amounts to 3 μm. The resulting DBP is expected to reach 1.2, indicating that a pulse can be delayed roughly by its pulse width. Absorption saturation at high excitation intensity may further increase the penetration depth, and thereby the DBP.